\documentclass{article}
\usepackage{spconf,amsmath,graphicx}
\usepackage{bbm}

\usepackage[normalem]{ulem}

\usepackage[lined, ruled]{algorithm2e}
\SetAlFnt{\small}
\SetAlCapFnt{\small}
\SetAlCapNameFnt{\small}
\SetAlCapHSkip{0pt}
\IncMargin{-\parindent}

\newtheorem{thm}{Theorem}[section]

\usepackage{tikz}

\DeclareMathOperator*{\argmax}{argmax} % no space, limits underneath in displays
\title{Fast Monte Carlo Dropout and Error Correction for Radio Transmitter Classification}
\name{Liangping Ma ~~ John Kaewell}
\address{InterDigital, San Diego, CA}
\begin{document}
%\ninept
%
\maketitle

\begin{abstract}
Monte Carlo dropout may effectively capture model uncertainty in deep learning, where a measure of uncertainty is obtained by using multiple instances of dropout at test time. However, Monte Carlo dropout is applied across the whole network and thus significantly increases the computational complexity, proportional to the number of instances. To reduce the computational complexity, at test time we enable dropout layers only near the output of the neural network and reuse the computation from prior layers while keeping, if any, other dropout layers disabled. Additionally, we leverage the side information about the ideal distributions for various input samples to do `error correction' on the predictions. We apply these techniques to the radio frequency (RF) transmitter classification problem and show that the proposed algorithm is able to provide better prediction uncertainty than the simple ensemble average algorithm and can be used to effectively identify transmitters that are not in the training data set while correctly classifying transmitters it has been trained on.  

%optimized by stochastic variational inference (SVI). We apply this to radio signal classification where the classifier needs to flag radio signals that do not belong to any of the known transmitters besides classifying signals from known transmitters. The experimental results show that the proposed algorithm is approximately Pareto optimal compared to other schemes.

\end{abstract}
\section{Introduction}
\label{sec:intro}
Dropout \cite{dropout14} was proposed as a means of regularization. The idea is to randomly drop units in a neural network during training time and then use the non-dropped trained neural network at test time. Monte Carlo dropout \cite{Gal16} uses dropout not only during training time but also at test time. For an input sample at test time, different instances of dropout produce different (deterministic) neural networks, which make different predictions, which in turn can be used to form measures on prediction uncertainty. However, to have a good measure on the prediction uncertainty, many instances of dropout are needed, and an input sample needs to pass through the neural network many times, leading to an increase in the computational complexity. To address this issue, at test time we enable one or more dropout layers only near the output of a neural network and reuse the outputs from earlier layers. This way, the computational complexity can be significantly reduced without affecting regularization.

As another contribution, we improve upon individual predictions by leveraging the side information that we have on the predictions. For a sample that belongs to one of the classes in the training dataset, we define the ideal output from the softmax layer as a discrete probability distribution whose peak is equal to one and corresponds to the correct class and whose other entries are all zero. We utilize this side information to correct the prediction: enlarge the peak to one and reduce other entries to zero.    

We propose an algorithm that uses the two techniques described above and apply it to the radio frequency (RF) transmitter classification problem~\cite{Oshea18}\cite{Merchant18}, which is motivated for tasks such as network security and RF interference mitigation. The present work differs from prior RF fingerprint work in that a prediction uncertainty measure is produced and used to identify potential RF transmitters on which the classifier is trained. We show that the proposed algorithm can 
effectively identify transmitters that were not part of the training dataset, while correctly classifying transmitters on which it has been trained.

The remainder of this paper is organized as follows. We present our main contributions in sections \ref{sec_fast} and \ref{sec_peak}. In section \ref{sec_exp} we show the experimental results followed by the conclusion in section \ref{sec_con}.

\section{Fast Monte Carlo dropout}\label{sec_fast}
\begin{figure}[htp]
\vspace{0.0in}  % signals.eps
  \includegraphics[width=2.8in]{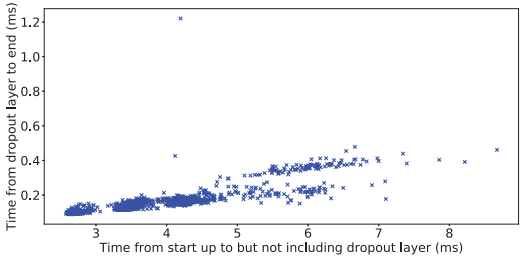}
\caption{The execution time for the post-dropout portion vs. the execution time for the pre-dropout portion for a 18-layer neural network on a GPU.}
\label{fig_speed}
\end{figure}

The direct approach to applying Monte Carlo dropout at test time in general is to pass the input sample through the entire neural network multiple times in order to generate multiple predictions and thereby incurring complexity proportional to the number of instances. To tackle this issue, at test time we enable one or more dropout layers near the output of the network while disabling, if any, other dropout layers, and reuse the computation up to the first enabled dropout layer. Specifically, we save the input to the first enabled dropout layer in the first pass, and in the remaining passes, we start from the saved input instead of recomputing it. This way, we can significantly reduce the computational complexity and memory usage (e.g., the dropout may be applied to a fully connected linear layer with smaller dimensions than the input sample, resulting in reduced storage for the second pass and so on), which may translate into savings in power consumption and time. The reduction depends on the network architecture. Generally, deeper networks, which tend to have dropout layers near the output, benefit more from this technique because there is greater reuse of the first inference. Since during training we use the same neural network (including dropout layers) as Monte Carlo dropout, \textit{our approach does not affect regularization.} We look at the savings on time for a particular neural network that we will use for the RF signal classification problem in later sections. The network consists of 16 convolutional layers (each followed by batch normalization\cite{Ioffe15}) with ResNet structure\cite{He16}, an average pooling layer, and a fully connected layer where dropout is applied and which feeds the softmax function. We measure the execution time for two parts of the computational graph: (i) the post-dropout portion (including dropout), and (ii) the pre-dropout portion. We run the experiment on a Nvidia GTX 1080 GPU for 1600 passes with batch size of 256 and plot the result in Fig.~\ref{fig_speed}. On average, the time of (ii) is about 25 times as large as the time of (i), meaning a 24x speed up for the second pass and onward.

\section{Prediction by ensemble average with error correction}\label{sec_peak}

First, for brevity, we define a \textit{known transmitter} to be a transmitter on which the classifier is trained, and a signal from which is called a \textit{known signal}. We define an \textit{unknown transmitter} to be one whose signals (called \textit{unknown signals}) are not included in the training dataset.

We then present the simple ensemble average algorithm suggested in \cite{Gal16}, and the error correction scheme which is incorporated into the simple ensemble average algorithm to form the proposed algorithm, followed by a theoretical analysis.  

\subsection{Simple Ensemble Average Algorithm}
We view the softmax output as an imprecise estimate of the previously defined ideal distribution. To have a better estimate, we can average many softmax outputs. However, since the noise is not zero mean, the averaging has the limitation of only producing a biased estimate. We call this algorithm the simple ensemble average algorithm, with details in Algorithm \ref{alg_avg}. At Line 3, the peak of the ensemble averaged distribution $t$ is compared with a threshold $\lambda$ to decide whether the input is an unknown/random signal or a known signal. If $t < \lambda$, the signal is considered as an unknown signal or random signal and the category corresponding to the peak is the predicted category. Otherwise, the signal is considered as known.

%There is a tradeoff between the probability of making a prediction and the accuracy of making a prediction for a signal from a known transmitter. Fig.~\ref{fig_thresh_all} shows how the accuracy changes as a function of the threshold for each of the three types of signals.

\IncMargin{1em}
\LinesNumbered
\begin{algorithm}\label{alg_avg}
\caption{Simple ensemble average}
%%\tcp{This function runs periodically every $\tau$ seconds. Default value for $\tau$ is 0.1.}
  \KwIn{$K$ distributions $v_{k}$, where $k=1, \dots, K$, obtained from \emph{Fast Monte Carlo dropout}; and threshold $\lambda$}
  \KwOut{known and category, or others}
%%  \tcp{$w$ is the window size. Default value of $w$ is 10. Replace the oldest with new one}
  $s = \sum_{k=1}^K v_{k}/K $ ~~//element-wise averaging\;  
  $t = \max_i s_i$ ~~// maximum entry of $s$\;
  \If{$t < \lambda$}
  {
  return others   /* unknown or random */ \;
  }
  return known, category = $\argmax_i s_i$ \;
\end{algorithm}
\DecMargin{1em}

%\begin{figure}[htp]  % 2019-02-17-02-21-05_Bayes-v2f2_thresh_all
%\vspace{0.0in}
%  \includegraphics[width=2.9in]{Figs/2019-08-19-17-57-55_Bayes-v2f2_thresh_all.eps}   % from file resnetRFFP-2-Lora-Bayes-v2f2.ipynb
%\caption{The ensemble average algorithm: the accuracy for a random signal (blue line with dots), accuracy for a signal from an unknown transmitter (red line with crosses), and overall accuracy for a signal from a known transmitter (green line with solid circles).}
%\label{fig_thresh_all}
%\end{figure}

\subsection{Error Correction}
The output of softmax is a vector, which we think of as a distribution giving the probability that the input sample belongs to each of the classes. For an input sample that belongs to one of the classes, if the prediction is correct, the probability mass of the correct class will be the peak of the probability vector, but may be less than 1. The \textit{ideal distribution} is a vector with probability 1 for the correct class and 0 for the other classes. This distribution is ideal in the sense that if the input class were known, then the probability of all other classes would be zero. We leverage this side information to `correct' the softmax output. Specifically, let the softmax output be $v_k$, and and the corrected be $\hat{v}_{k}$, where $k=1,\dots, N$ and $N$ is the number of instances of the neural network with dropout. We consider two thresholds  $\beta_1$ and $\beta_2$ such that $\beta_1 < \beta_2$. If the peak of $v_k$, i.e., $\max_i v_{k,i}$ is large enough ($> \beta_2$), we enlarge the peak to 1 and shrink the other probabilities to 0, as shown below
\begin{equation}
\hat{v}_k = \left \{
\begin{array}{lr}
(0, \dots, 1, \dots, 0), & \mathrm{if~}\max_i v_{k,i} \geq \beta_2\\
v_k, & \mathrm{if~} \beta_1 \leq \max_i v_{k,i} < \beta_2 \\
(\frac{1}{C}, \dots, \frac{1}{C}), & \mathrm{if~}\max_i v_{k,i} < \beta_1 
\end{array} 
\right. \label{eq_thresholds}
\end{equation}
where $v_{k, i}$ is the probability mass for class $i$ in vector $v_k$, $i=0,2, \dots, C-1$, and the $1$ in the first row is at location $\argmax_i v_{k,i}$. An example is illustrated in Fig. 2 (a), where category $c$ is the correct class. For an input sample that does not belong to one of the classes, the ideal output will be a uniform distribution. Therefore, if $\max_i v_{k,i}$ is small enough ($< \beta_1$), which may indicate high uncertainty, we `correct' the softmax output to a uniform distribution. If $\max_i v_{k,i}$ is between the two thresholds, we do not change $v_k$. Clearly this procedure does not improve the correctness of predictions of individual neural networks. We investigate if the procedure applied to an ensemble of neural networks as detailed in Algorithm \ref{alg_1} could be used to improve the uncertainty measure of a prediction compared to the ensemble average algorithm.

If we set $\beta_1=0$ and $\beta_2=1$, then it follows from (\ref{eq_thresholds}) that $\hat{v}_k = v_k$ always holds, the same as Algorithm \ref{alg_avg}. In other words, Algorithm \ref{alg_1} includes Algorithm \ref{alg_avg} as a special case. If we properly choose the values for $\beta_1$ and $\beta_2$, the performance of Algorithm \ref{alg_1} will be at least as good as that of Algorithm \ref{alg_avg}. 
   
\begin{figure}[htp]
\vspace{0.0in}
  \includegraphics[width=2.5in]{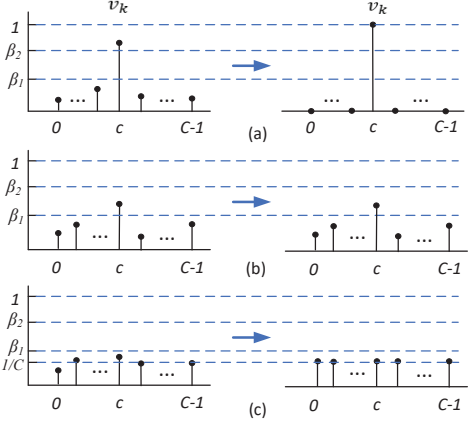}
\caption{Error correction on the softmax output, where $c$ is the correct category and the peak is assumed to be at class $c$.}
\label{fig_peak_mass}
\end{figure}

Note that if the correct class is $c$ and the prediction is correct, i.e., $\argmax_i v_{k,i}=c$, the correction procedure for the case $\max_i v_{k,i} < \beta_1$ may bring down the probability of class $c$, as shown in Fig. 2 (c). The remaining case $\beta_1 \leq \max_i v_{k,i} < \beta_2$ does not affect the probability of class $c$, as shown in Fig. 2 (b). Also, the prediction from a neural network may be wrong. Thus, we need to consider all cases to evaluate the benefit of the propose procedure.

We now give the conditions under which the error correction procedure provides an overall gain in predicting an input sample that belongs to one of the categories for which an classifier is designed. Let the set of deterministic DNNs that make a right prediction be $\mathcal{R}=\{k | v_{k,c} > v_{k,i}, k=1, 2, \dots, N, \forall i \neq c\}$, and the complementary set be $\mathcal{W}$. First consider a network $k \in \mathcal{R}$. With Algorithm \ref{alg_1}, the correct category $c$ gains all the masses from the other categories equal to $1-v_{k,c}$ if $v_{k,c} > \beta_2$, as shown in Fig.~\ref{fig_peak_mass}(a). Category $c$ loses probability mass of $v_{k,c} - 1/C$ if $v_{k,c} < \beta_1$ as shown in Fig.~\ref{fig_peak_mass}(c). Next consider a network $k \in \mathcal{W}$. The correct category $c$ loses all its mass $v_{k,c}$ if $\max_i v_{k,i} > \beta_2$ and gains mass $1/C - v_{k,c}$ if $\max_i v_{k,i} < \beta_1$. Overall, the mass (normalized by $K$) for category $c$ is $p_{2}=(1/K)(\sum_k v_{k, c} + \sum_{k \in \mathcal{R}} ((1- v_{k,c})\mathbbm{1}(v_{k,c} > \beta_2)-\sum_{k \in \mathcal{R}}(v_{k,c} - 1/C)\mathbbm{1}(v_{k,c} < \beta_1) -\sum_{k \in \mathcal{W}}v_{k,c}\mathbbm{1}(\max_i{v_{k,i}} > \beta_2)+ \sum_{k \in \mathcal{W}} ((1/C - v_{k,c})\mathbbm{1}(\max_i{v_{k,i}} < \beta_1))$, where $\mathbbm{1}()$ is the indicator function, equal to 1 when the argument is true and 0 otherwise. As a comparison, the ensemble average algorithm (Algorithm \ref{alg_avg}) yields a mass for category $c$ of $p_{1} = (1/K)(\sum_k v_{k, c})$. The difference $\Delta_c := p_{\mathrm{2}} - p_{\mathrm{1}}$. Then,
\begin{eqnarray}
K \Delta_c =\sum_{k \in \mathcal{R}} (( 1-v_{k,c})\mathbbm{1}(v_{k,c} > \beta_2)-(v_{k,c} - 1/C)\mathbbm{1}(v_{k,c} < \beta_1)) & \nonumber \\
- \sum_{k \in \mathcal{W}} (v_{k,c}\mathbbm{1}(\max_i{v_{k,i}} > \beta_2)-(1/C-v_{k,c}) \mathbbm{1}(\max_i{v_{k,i}} < \beta_1)) & \nonumber
%\label{eq_diff}
\end{eqnarray}
To simplify the result, we define conditional expectations for $v_{k,c}$: $\bar{v}_c^{\mathcal{R}, \beta_2^+}$ for the expectation conditioned on $k \in \mathcal{R}$ and $v_{k,c}> \beta_2$, $\bar{v}_c^{\mathcal{R}, \beta_1^-}$ for $k \in \mathcal{R}$ and $v_{k,c}< \beta_1$, and similar notations for $k \in \mathcal{W}$. We also define conditional distribution functions: $F_c^{\mathcal{R}}(\beta_2)=\mathrm{Prob}[v_{k,c} \leq \beta_2|k\in \mathcal{R}]$, $F_m^{\mathcal{W}}(\beta_2)=\mathrm{Prob}[\max_i v_{k,i} \leq \beta_2|k\in \mathcal{W}]$. Then, as $K \to \infty$
\begin{eqnarray}
\Delta_c =\alpha ((1-\bar{v}_c^{\mathcal{R}, \beta_2^+})(1-F_c^{\mathcal{R}}(\beta_2)) - (\bar{v}_c^{\mathcal{R}, \beta_1^-}-1/C)F_c^{\mathcal{R}}(\beta_1)) \nonumber \\
- (1-\alpha)(\bar{v}_c^{\mathcal{W}, \beta_2^+}(1-F_m^{\mathcal{W}}(\beta_2)) -(1/C - \bar{v}_c^{\mathcal{W}, \beta_1^-}) F_m^{\mathcal{W}}(\beta_1) ) \label{eq_diff}
\end{eqnarray}
where $\alpha$ is the probability that an instance of the neural network makes a correct prediction. If the above is greater than 0, the probability mass for the correct class will increase. To further simplify the analysis, suppose that the other classes have equal probability mass with Algorithm \ref{alg_1}. Since the total probability mass is equal to 1, any of the incorrect class will have a reduced probability mass if $\Delta_c >0$. This is summarized below.
\begin{thm}
When $K \to \infty$, if $\alpha ((1-\bar{v}_c^{\mathcal{R}, \beta_2^+})(1-F_c^{\mathcal{R}}(\beta_2)) - (\bar{v}_c^{\mathcal{R}, \beta_1^-}-1/C)F_c^{\mathcal{R}}(\beta_1)) > (1-\alpha)(\bar{v}_c^{\mathcal{W}, \beta_2^+}(1-F_m^{\mathcal{W}}(\beta_2)) -(1/C - \bar{v}_c^{\mathcal{W}, \beta_1^-}) F_m^{\mathcal{W}}(\beta_1) )$, Algorithm \ref{alg_1} provides a more accurate probability distribution for the correct category than Algorithm \ref{alg_avg}.
\label{theorem_better}
\end{thm}
The condition in Theorem \ref{theorem_better} is easily met if $\alpha$ is close to 1 and $\beta_1$ is relatively small.

\IncMargin{1em}
\LinesNumbered
\begin{algorithm}\label{alg_1}
\caption{Ensemble average with error correction}
%%\tcp{This function runs periodically every $\tau$ seconds. Default value for $\tau$ is 0.1.}
  \KwIn{$K$ distributions $v_{k}$, where $k=1, \dots, K$, obtained from \emph{Fast Monte Carlo dropout}; and $\beta_1$, $\beta_2$, $\lambda$}
  \KwOut{known and category, or others}
%%  \tcp{$w$ is the window size. Default value of $w$ is 10. Replace the oldest with new one}
  Compute $\hat{v}_k$ according to Eq. (\ref{eq_thresholds}) \;
  $s = \sum_{k=1}^K \hat{v}_{k}/K $ ~~//element-wise averaging\;  
  $t = \max_i s_i$ ~~// maximum entry of $s$\;
  \If{$t < \lambda$}
  {
  return others   /* unknown or random */ \;
  }
  return known, category = $\argmax_i s_i$ \;
\end{algorithm}
\DecMargin{1em}

\subsection{Performance for unknown or random signals}
For these two cases, the \emph{ideal distribution} is $f_i=1/C$, where $i=0, ..., C-1$, which represents maximum uncertainty in prediction. Write the softmax output $v_{k,i} = 1/C + \epsilon_i$ for $i=0, ..., C-1$ and instance $k$, where random variables $\epsilon_i$ satisfy $\sum_i \epsilon_i =0$ and $ - 1/C \leq \epsilon_i \leq 1 - 1/C $. The last constraint ensures $v_{k,i}$ to be non-negative. Assume that $\epsilon_i$'s are identically distributed. Then the ensemble averaging algorithm -- Algorithm \ref{alg_avg} -- produces a uniform distribution as $K \to \infty$. From (\ref{eq_thresholds}), there is no preference to any category under Algorithm \ref{alg_1}. Thus, by symmetry, as $K \to \infty$, Algorithm \ref{alg_1} also produces a uniform distribution. Therefore, both algorithms have the same performance for unknown and random signals. 

\section{Experimental Result}\label{sec_exp}

The LoRa radio \cite{Lora} is a commercial long-range (e.g., 10km) and low-power radio characterized by chirp spread spectrum modulation and intended for the Internet of Things (IoT) applications. The signals are captured using USRP X310 with an oversampling factor of 4. The channel bandwidth is 1MHz at a carrier frequency of 902.3MHz. 1000 signals are captured for each of the 48 LoRa transmitters, 44 of which are the known categories and the remaining 4 the unknown. Each signal consists of 500 complex samples at the beginning of the packet including a ramp up portion and 500 complex samples at the end of the packet including a ramp down portion with 1000 IQ samples in total. Random shift in time is used for data augmentation.

\begin{figure}[htp]
\vspace{-0.5in}
  \includegraphics[width=2.8in]{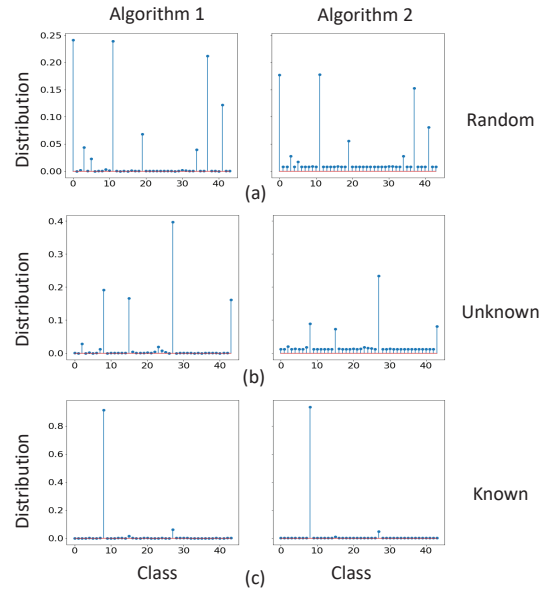}
\caption{Distributions resulting from Algorithm 1 (left column) and Algorithm 2 (right column) for (a) a random signal (Gaussian), (b) an unknown signal, and (c) a known signal.}
\label{fig_dist}
\end{figure}

\begin{figure}[htp]
\vspace{-0.5in}
  \includegraphics[width=2.7in]{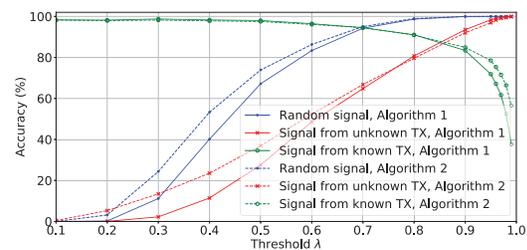}
\caption{The prediction accuracy by Algorithm 1 (solid lines) and by the proposed Algorithm 2(dashed lines).}
\label{fig_compare}
\end{figure}

Figure \ref{fig_dist} shows the distributions resulting from Algorithm \ref{alg_avg} (the left column) and Algorithm \ref{alg_1} (right column) for Gaussian random signals (top), unknown signals (middle), and known signals (bottom), where $K=500, \beta_1 =0.50, \beta_2=0.92$. The same input data and noise realization are used for each row for a fair comparison. The noise is generated with SNR$=20$dB. We see that the resulting distributions from Algorithm \ref{alg_1} are closer to the respective ideal distributions. Further comparison is available in Fig.~\ref{fig_compare}. Algorithm \ref{alg_1} provides the same accuracy or better accuracy than Algorithm \ref{alg_avg} for all three types of signals, except for $\lambda \in [0.75, 0.95]$, where Algorithm \ref{alg_1} is slightly worse for the unknown signals. 

\section{Conclusion}\label{sec_con}
We proposed a speed-up technique and an error correction technique for the Monte Carlo dropout algorithm and apply them to the RF transmitter classification problem with unknown transmitters. We gave a theoretical analysis. The effectiveness was shown through experimental results.  
\subsubsection*{Acknowledgements}
The authors would like to thank Dr. Philip Pietraski of InterDigital for insightful comments, and Sudhir Pattar and Ishan Sethi (as an intern at InterDigital) for collecting the radio signal data for the LoRa transmitters.

\bibliographystyle{IEEEbib}
\bibliography{refs}

\end{document}